\documentclass[12pt]{iopart}

\usepackage{siunitx}
\usepackage{xcolor}
\usepackage{graphicx}
\usepackage{soul}
\usepackage[font={small}]{caption}
\usepackage{comment}

\usepackage{float}
\begin{document}
	\title[Proton beams accelerated from an overdense gas jet]{Spectrally peaked proton beams shock accelerated from an optically shaped overdense gas jet by a near-infrared laser}

	\author{G S Hicks$^1$, O C Ettlinger$^1$, M Borghesi$^2$, D C Carroll$^3$, R J Clarke$^3$, E J Ditter$^1$, T P Frazer$^4$, R J Gray$^4$, A McIlvenny$^2$, P McKenna$^4$, C A J Palmer$^5$\footnote{Present address: Centre for Plasma Physics, Queen's University Belfast, Belfast, BT7 1NN, UK}, L. Willingale$^6$ and Z Najmudin$^1$}
	
	\address{$^1$ The John Adams Institute for Accelerator Science, Blackett Laboratory, Imperial College London, London, SW7 2BW, UK}
	\address{$^2$ Centre for Plasma Physics, Queen's University Belfast, Belfast, BT7 1NN, UK}
	\address{$^3$ Central Laser Facility, STFC Rutherford Appleton Laboratory,  Oxfordshire, OX11 0QX, UK}
	\address{$^4$ SUPA, Department of Physics, University of Strathclyde, Glasgow, G4 0NG, UK}
	\address{$^5$ Physics Department, Lancaster University, Lancaster, LA1 4YB, UK}
	\address{$^6$ University of Michigan, 2200 Bonisteel Boulevard, Ann Arbor, Michigan 48109, USA}
	\ead{g.hicks11@imperial.ac.uk}
	
	\begin{abstract}
		We report on the generation of impurity-free proton beams from an overdense gas jet driven by a near-infrared laser ($\lambda_L=1.053$\,\si{\micro\metre}).
		The gas profile was shaped prior to the interaction using a controlled prepulse. 
		Without this optical shaping, a \SI[separate-uncertainty = true]{30(4)}{\nano\coulomb\per\steradian} thermal spectrum was detected \emph{transversely} to the laser propagation direction with a high energy \SI[separate-uncertainty = true]{8.27(7)}{MeV}, narrow energy spread (\SI[separate-uncertainty = true]{6(2)}{\%}) bunch containing \SI[separate-uncertainty = true]{45(7)}{\pico\coulomb\per\steradian}. In contrast, with optical shaping the radial component was not detected and instead forward going protons were detected with energy \SI[separate-uncertainty = true]{1.32(2)}{MeV}, \SI[separate-uncertainty = true]{12.9(3)}{\%} energy spread, and charge \SI[separate-uncertainty = true]{400(30)}{\pico\coulomb\per\steradian}. Both the forward going and radial narrow energy spread features are indicative of collisionless shock acceleration of the protons.
	\end{abstract}
	
	%\noindent{\it Keywords: ion acceleration, near-critical density, high repetition-rate}

	%=================================================================================
	\section{Introduction}\label{sec:introduction}
	%=================================================================================
	%textwidth in cm: \printinunitsof{cm}\prntlen{\textwidth}
	Laser-plasma ion accelerators have the capability to be much smaller than conventional accelerators since they can support much stronger accelerating fields \cite{Borghesi2007,Daido2012a,Macchi2013b,McKenna2013,Macchi2017}. Compact sources of energetic ions would be of interest for a number of applications including radiography, medical isotope production, radiobiology and hadron therapy \cite{Wilson1946,Borghesi2002,Bulanov2002,Fritzler2003,Linz2007,Murakami2008,Doria2012}.
%Foams targets however present a number of challenges when implemented experimentally, including low rep-rate, density inhomogeneities and potential multi-species ion beams

	High-density gas jets are attractive targets for laser-driven ion acceleration since they can provide homogeneous, easily replenishable near-critical density targets suitable for driving high-speed collisionless shocks. These shocks can be driven directly by the radiation pressure of the laser \cite{Palmer2011, Robinson2009b}, in which case the acceleration is often called hole-boring radiation pressure acceleration (HB-RPA), or by thermal pressure due to the absorbed laser energy, which is usually called collisionless shock acceleration (CSA) \cite{Denavit1992, Silva2004a}.
	In either case, the action of the laser launches an electrostatic collisionless shock through the plasma at a speed $v_{sh}$ \cite{Denavit1992, Silva2004a}. Upstream ions can reflect from this shock at velocities up to $v=2v_{sh}$. If $v_{sh}$ is near constant or the reflection time is short, the accelerated ions have a narrow energy spread \cite{Palmer2011}. This is highly desirable for many applications as it negates the need for complex energy filtering, increasing the simplicity of using these beams. 
	
	The shock velocity is faster for targets of lower density, $n$, since the velocity is ultimately limited by the amount of upstream material that acts to impede its motion. Hence, the energy of the accelerated ions, $\mathcal{E}_{\it HB}$, scales as $\mathcal{E}_{\it HB}\propto {1}/{n}$ \cite{Robinson2009b}, and using a lower density plasma increases the maximum observed proton energies. However, the target must still have a plasma density exceeding the critical density, as below this density both radiation pressure and laser absorption become ineffectual.
	
	The requirement for near-critical density targets meant that shock acceleration in the laser direction was first demonstrated using $\lambda \approx \SI{10}{\micro m}$ wavelength CO$_2$ lasers \cite{Palmer2011, Haberberger2011}. The critical  plasma electron density for $\lambda\approx\SI{10}{\micro\metre}$ is $n_c\approx\SI{1e25}{\per\metre\cubed}$. This is easily achievable with a gas jet, since it is close to the atomic density of air at atmospheric pressure, $n_{\mathrm{air}}\approx \SI{2.5e25}{\per\metre\cubed}$. Using CO$_2$ lasers, shock acceleration was used to accelerate protons to 1.1\,\si{\mega\electronvolt} at Brookhaven National Laboratory (BNL) \cite{Palmer2011} and to 22\,\si{\mega\electronvolt} at University of California Los Angeles \cite{Haberberger2011}. The BNL experiment showed that the acceleration is strongly dependent on laser intensity. 
	The brightest CO$_2$ lasers currently have  peak vacuum intensities of $I\approx\SI{e22}{Wm^{-2}}$ \cite{Polyanskiy2020}. By contrast, near-infrared (NIR) laser systems ($\lambda=\SIrange[range-units=single]{0.8}{1.1}{\micro\metre}$) can reach intensities up to or exceeding $I=\SI{e27}{Wm^{-2}}$ \cite{Danson2019}. However, the critical electron plasma density for a typical NIR laser of wavelength $\lambda_L=\SI{1.053}{\micro\metre}$ is $n_c=\SI{1.0e27}{\per\metre\cubed}$. This is much less than the  electron density of a typical solid target; e.g. for formvar plastic $n_{\mathrm{formvar}}=\SI{4e29}{\per\metre\cubed}$.
	Hence, producing near-critical density targets for optical or NIR lasers is challenging, requiring either an expanded, initially solid material or a very high-density gas.
	
	Foam targets are typically a loose aggregation of gas or voids inside a plastic structure which can be expanded to create near-critical density targets \cite{Willingale2009,Bin2015}. Each foam must be aligned to the laser individually making the targets low repetition-rate. A multi-species ion beam will be generated since the foam is plastic and the required metal-mount can cause debris which can damage nearby components such as beam transport magnets. Since the foams are formed of regions of high-density with voids, they are initially strongly inhomogeneous. This may alter the interaction and may require homogenisation with a pre-heating laser which adds complication to their use.
	
	By contrast, gas targets are potentially high repetition-rate, usually only limited by the rate that the vacuum pumps remove residual gas. The choice of gas species is flexible, can easily be single-species and can be changed rapidly. A disadvantage of gas targets compared to foams or solid targets is that the density profile is not well suited for ion acceleration. Typical gas profiles are Gaussian with long initial density scale-lengths compared to much sharper solid interfaces. Laser filamentation can occur in the low-density plasma in the rising-edge of the gas profile, which leads to depletion of the laser energy before the laser reaches the plasma critical surface, as discussed later in this paper.
	
	However, it is possible to use a controlled prepulse to shape a gas jet. This technique has been shown to form short scale-length features in gas jets on the 10.6\,\si{\micro\metre} wavelength laser system at BNL \cite{Tresca2015,Dover2016a}.
	In the work by Tresca {\it{et al}},	a low intensity laser prepulse ($I \sim \SI{e18}{Wm^{-2}}$) was followed by a high intensity pulse ($I= \SI{2.5e20}{Wm^{-2}}$). The prepulse was used to drive a blast wave inside a gas target to steepen the gas profile. This gas profile steepening is advantageous for two reasons. Firstly, the blast wave sweeps up gas to create a short scale-length, high-density region. This high-density region has a density, $\rho = \rho_0 ({\gamma+1})/({\gamma-1})$ \cite{Sedov1959}, where $\rho_0$ is the initial density and $\gamma$ is the ratio of specific heats of the gas. For hydrogen, $\gamma=1.41$, so $\rho=5.88\, \rho_0$. Secondly, due to this gas accumulation, the region behind the density spike is void of gas. This enables the high intensity pulse to interact with the critical density surface of the plasma without being distorted and depleted by the underdense plasma. This method was used in \cite{Tresca2015} to accelerate He$^{2+}$ ions up to 2\,\si{\mega\electronvolt}. 
	
	A second issue with gas targets is that reaching the electron densities of $10^{27}\,\si{\per\metre\cubed}$ required for use with optical lasers is technically challenging and requires very high backing pressures of order 500\,\si{\bar}.
	Gas targets have been used in an experiment to generate proton beams at the PICO2000 facility at LULI \cite{Puyuelo-Valdes2019} with an intensity of $I=\SI{4e23}{Wm^{-2}}$ and at the Titan laser, LLNL \cite{Chen2017a} with an intensity of $I=\SI{2e23}{Wm^{-2}}$. Though accelerated ions were observed in these experiments and were attributed to CSA, the high-charge feature peaked at high energy, that is characteristic of shock acceleration, was not observed. Also, the optical shaping laser pulse energy and delay were not controlled. Gas targets have also been used at the Naval Research Laboratory with an intensity of $I=\SI{1e20}{Wm^{-2}}$. In this work two shocks were launched from the front and rear of a hydrogen gas jet. Up to \SI{2}{\mega\electronvolt} protons were accelerated from a \SI{75}{\micro\metre} thick gas ``foil" generated where the shocks collide. The acceleration mechanism was attributed to magnetic vortex acceleration \cite{Helle2016}. Recently, a study by J.-R. Marques {\it{et al.}} have shown using simulations that two parallel nanosecond pulse-duration beams can generate a gas foil suitable for proton acceleration by a perpendicularly propagating driver beam \cite{Marques2020}. 
	
	In this paper, we report on experiments to implement optical shaping on high-density gas jets. We measure the characteristics of shock acceleration from the interaction of intense NIR lasers with density-steepened near-critical density targets for the first time.
	
	\section{Experiment}
	%=================================================================================
	The work was performed using the Vulcan Petawatt laser at the Rutherford Appleton Laboratory.
	A prepulse was created using a beam with mean energy,  $E_{PP}=207\pm7\,\si{\milli\joule}$, mean pulse-length $\tau_{PP}=4.2\pm0.2$\,ns and wavelength $\lambda_{PP}=1.053$\,\si{\micro\metre}, where the mean energy and pulselength are calculated from all the shots taken under the conditions presented in this paper and the errors presented refer to the standard error on the mean.
	The vertically polarised prepulse was focussed with an $f/10$ lens and aligned by obscuration on a \SI{250}{\micro\metre} wire.
	A diffraction limited focal spot of full-width at half-maximum (FWHM) \SI{10.1}{\micro\metre} gave a prepulse intensity of $I_{\mathrm{PP}}=\SI{4.2e17}{Wm^{-2}}$. The horizontally polarised main pulse was operated with mean on-target energy, $E_{MP}=207\pm5\,\si{\joule}$ and pulse length, $\tau_{MP}=0.61\pm0.05$\,ps and wavelength $\lambda_{MP}=1.053$\,\si{\micro\metre}, where the errors presented refer to the standard error on the mean. The focal spot was measured at low power and 38\% of the energy was found to be contained with a spot of FWHM size 5.4\,\si{\micro\metre}, which was assumed to be the same at high power. This gives a theoretical peak intensity in the main pulse of $I_{\mathrm{MP}}=\SI{7.4e24}{Wm^{-2}}$.
	
\begin{figure}[b]
	\centering
	\begin{minipage}{\textwidth}
	\begin{minipage}[t]{0.545\textwidth}
		\includegraphics[width=0.99\textwidth]{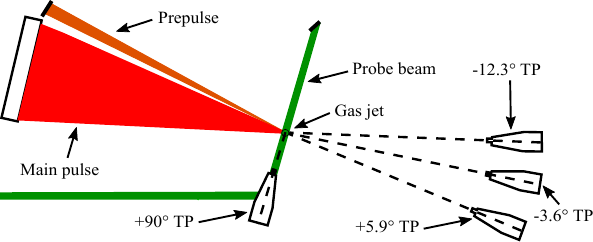}
		\caption[Layout]{Experimental layout: Gas nozzle located at the centre of the chamber where both main and prepulse focus. The interaction was transversely probed by a frequency-doubled pick-off of the main beam. 4 Thomson Parabola (TP) spectrometers at \ang{-11.1}, \ang{-3.3}, +\ang{5.9} and +\ang{90} with respect to the laser propagation axis diagnosed the accelerated ion beam. The +\ang{90} TP was positioned above the horizontal plane, pointing down at the interaction point.}
		\label{fig:layout}
	\end{minipage}
	\hspace{0.02\textwidth}
	\begin{minipage}[t]{0.42\textwidth}
		\includegraphics[width=0.99\textwidth]{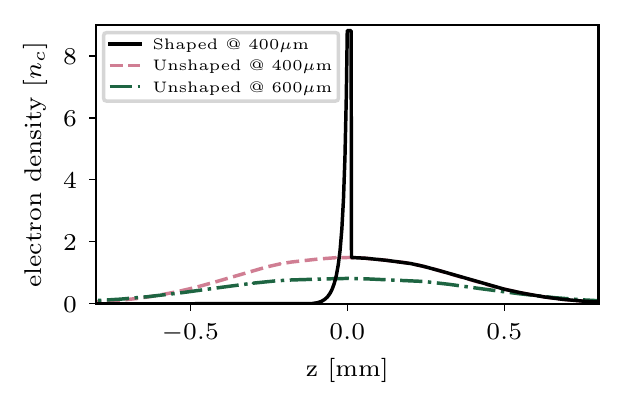}
		\caption[gas jet profiles]{Electron density profiles from an unshaped gas jet at heights of \SI{400}{\micro\metre} and \SI{600}{\micro\metre} above the nozzle from pre-experiment characterisations and from shaped gas at a height of \SI{400}{\micro\metre} as calculated from eq$\mathrm{\underline{n}}$ \ref{eq:sedov}. The densities correspond to critical densities for a laser wavelength of $\lambda=1.053$\,\si{\micro\metre}.}
		\label{fig:gas_jet_profiles}
	\end{minipage}
	\end{minipage}
	\end{figure}

	The experimental layout is shown in figure \ref{fig:layout}. The Petawatt laser was incident on a gas jet produced by forcing hydrogen gas through a subsonic gas nozzle with a \SI{0.5}{\milli\metre} diameter orifice. The prepulse was focussed onto the gas jet from a mirror positioned outside the $f/3$ cone of the main beam at an 18$^\circ$ relative angle. A pick-off from the main beam was frequency doubled to \SI{527}{nm} and used as a transverse probe beam.
	%An radiochromic film (RCF) stack was placed \SI{50}{mm} from the gas jet centred on the laser axis. The RCF stack consisting of layers of HD-V2 film followed by layers of EBT2 film, with metal filters placed in between these RCF layers to adjust the proton energy range interrogated.
	
\begin{figure}
	\centering
	\includegraphics[width=0.99\textwidth]{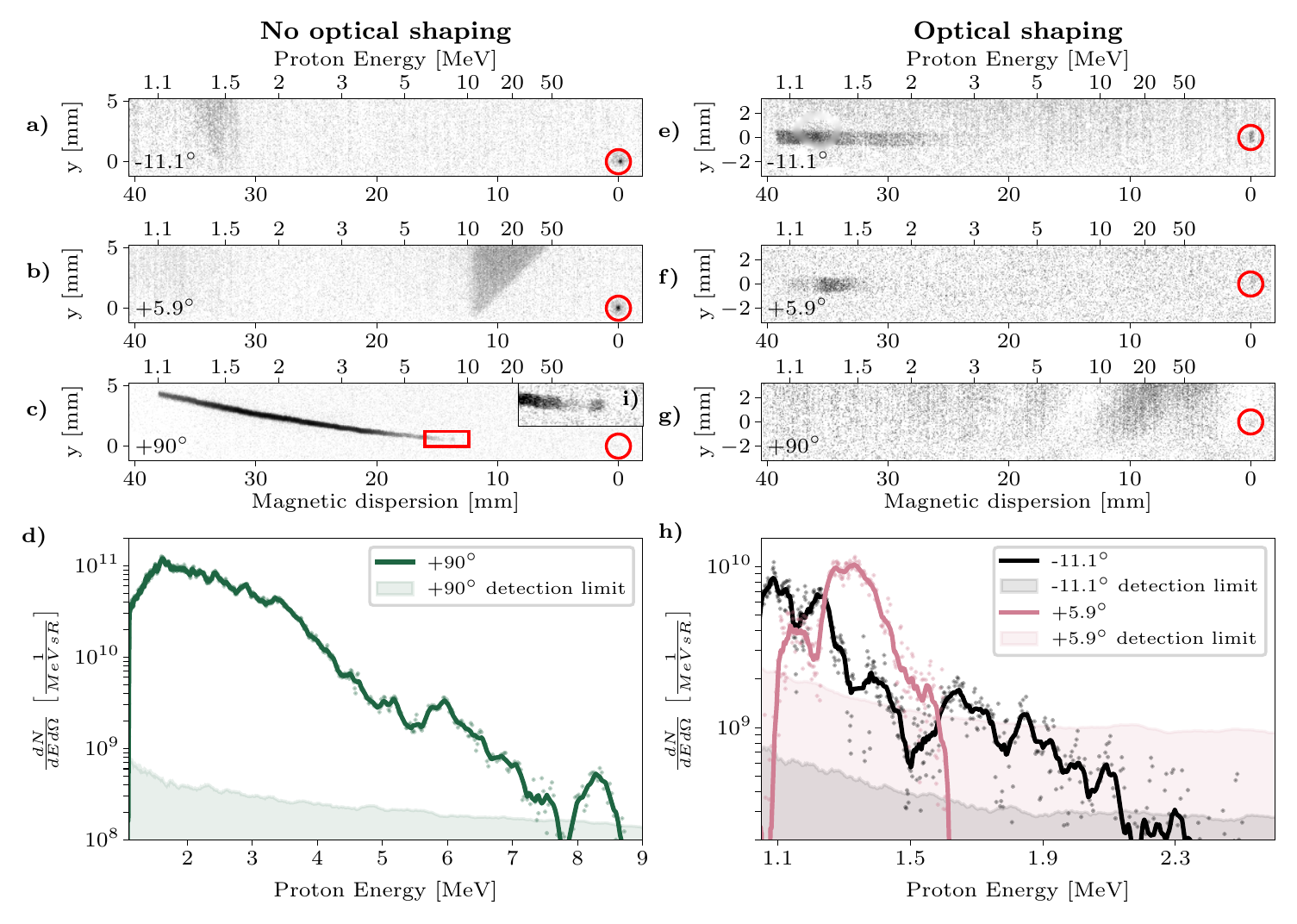}
	\caption[Thomson parabola (TP) spectrometer images and spectra]
	{
	Background-subtracted Thomson parabola image plate scans for shots:
	a-c) with no optical shaping at a height of \SI{400}{\micro\metre} above the nozzle and
	e-g) with optical shaping at a height of \SI{600}{\micro\metre} above the nozzle. Red circles indicate neutral points, and ions are deflected to the right
	The electric field of the TP spectrometer was turned off during the optical shaping shot and so the signal would be expected to be at the same y-position as the zero point.
	d) Spectrum for the +\ang{90} TP with no optical shaping;
	h) spectra for the \ang{-11.1} and +\ang{5.9} TPs with optical shaping.
	The inset i) shows the high energy peak at the end of the TP trace in c).
	The shaded regions in d) and h) are the 3$\sigma$ detection limits for each of the three extracted spectra. A 7 pixel moving average filter was applied to the spectra in d) and a 15 pixel moving average filter was applied to the spectra in h), the unfiltered individual data points are also plotted.
	The TPs in a-d used a pinhole of diameter 200\,\si{\micro\metre} (forward TPs) or 250\,\si{\micro\metre} (\ang{+90} TP) and the ions were additionally dispersed by electric field plates.
	The TPs in e-h used 300\,\si{\micro\metre} wide by 1.2\,mm long slits and the electric field plates were turned off.
	}
		\label{fig:TP_data}
	\end{figure}

	Three Thomson parabola (TP) spectrometers were placed at \ang{-11.1}, \ang{-3.3} and +\ang{5.9} with respect to the laser propagation axis.  These TPs are hereinafter reffered to as the``forward" TPs. An additional TP was placed at +\ang{90} to the laser axis above the horizontal plane, pointing down at the interaction point. Ions were dispersed within the spectrometers \cite{Carroll2006} by a \SI{50}{mm} long, \SI{0.6}{T} permanent magnet before propagating \SI{200}{mm} to the detector, BAS-TR image plate, which was scanned on a FUJIFILM FLA-5100 scanner. The proton flux was extracted using the calibration presented by Mancic {\it{et al}} \cite{Mancic2008}. Early in the experiment, and for the unshaped gas data presented later in figure \ref{fig:TP_data}, each TP sampled the proton beam with a pinhole of diameter 200\,\si{\micro\metre}{ (forward TPs) or 250\,}\si{\micro\metre} (\ang{+90} TP) and the ions were additionally dispersed by electric field plates. However, since the only ions to be accelerated were protons, the pinholes were changed for 300\,\si{\micro\metre} wide by 1.2\,mm long slits and the electric field plates were turned off for the majority of the shots. No signal was detected on the \ang{-3.3} TP for either the shaped or unshaped gas, so this data is not shown.
	%The RCF was scanned with using a Nikon LS-9000ED
	
	Initially, shots were taken into unshaped gas at varying heights above the nozzle at the maximum possible backing pressure of \SI{240}{bar}. This produced a gas jet with electron density of $1.5\,{n_c}$ at a height of \SI{400}{\micro\metre} above the nozzle and $0.82\,{n_c}$ at \SI{600}{\micro\metre} above the nozzle. The corresponding transverse gas profiles are shown in figure \ref{fig:gas_jet_profiles}. The main pulse was focussed to the centre of the gas jet. No protons were detected on any of the TPs when the laser was focussed \SI{400}{\micro\metre}. With  the laser was focussed \SI{600}{\micro\metre} no protons were detected on the forward TPs, as can be seen in figure \ref{fig:TP_data}\,(a),(b). However, protons were detected on two shots on the +\ang{90} TP. The raw image plate scan from one of these shots is  shown in figure \ref{fig:TP_data}(c). The detected protons had a thermal spectrum of charge \SI[separate-uncertainty = true]{30(4)}{\nano\coulomb\per\steradian} with a high energy bunch at \SI[separate-uncertainty = true]{8.27(7)}{MeV} of charge \SI[separate-uncertainty = true]{45(7)}{\pico\coulomb\per\steradian} separated from the thermal spectrum which falls into the noise above 7.7\,MeV. The high energy bunch had a \SI[separate-uncertainty = true]{6(2)}{\%} energy spread and is shown in the inset, figure \ref{fig:TP_data}(i). The direction of the protons is characteristic of the previously observed Coulomb explosion of a cylindrical underdense plasma column from which all of the electrons have been ponderomotively expelled \cite{Krushelnick1999}. It has also been observed previously that the energy of the expelled ions could be enhanced beyond the ponderomotive potential by shock acceleration of ions off the cylindrical expansion created by the Coulomb explosion, resulting in a density dependence on the energy gain \cite{Wei2004,Willingale2006,Singh2020}. Hence, the presence of a narrow energy spread feature beyond the thermal distribution, as observed in this shot, is a strong indication of the presence of shock acceleration.

	\begin{figure}
	\centering
	\begin{minipage}{\textwidth}
	\begin{minipage}[t]{0.55\textwidth}
	\includegraphics[height=4.5cm]{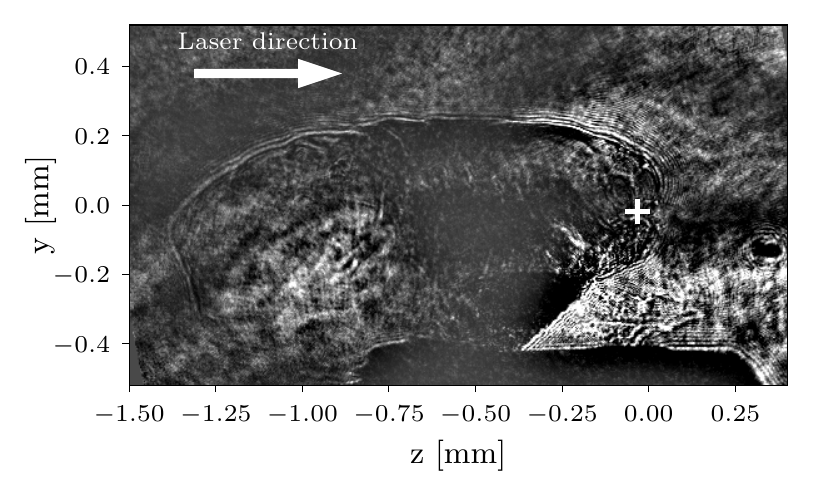}
	\caption{Shadowgraphy image of a blast wave \SI{6.08}{\nano\second} after the peak of the prepulse of energy \SI{219}{\milli\joule}. The arrow indicates the laser direction and the white cross indicates the main beam focus. The gas nozzle can be seen at the bottom of the image. A Gaussian high pass filter has been applied to the image to reduce non-uniformities in the backlighting beam.}
	\label{fig:probe}
	\end{minipage}
	\hspace{0.02\textwidth}
	\begin{minipage}[t]{0.43\textwidth}
	\centering
	\includegraphics[height=4.5cm]{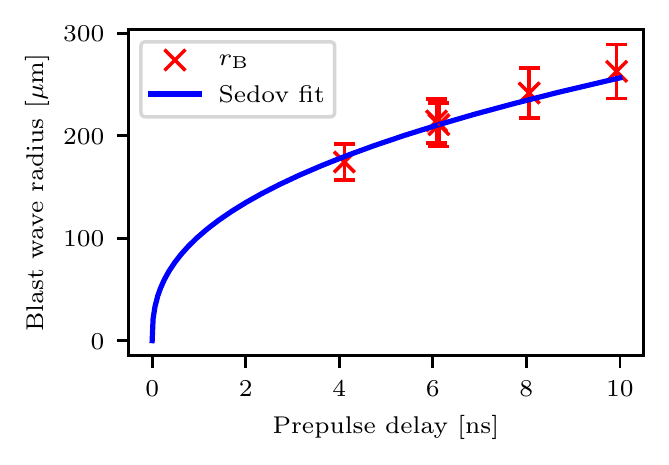}
	\caption{Blast wave radius, $r_\mathrm{B}$ as a function of time, $t$ as measured by shadowgraphy (crosses), and best fit with a Sedov expansion as per eq$^\mathrm{\underline{n}}$ \ref{eq:sedov} (line) where $E_{ab}=\SI{31}{\milli\joule}$. }
	\label{fig:blastwave_radius_v_time}
	\end{minipage}
	\end{minipage}
	\end{figure}
		
	A series of measurements were then taken with the low energy prepulse without the main pulse to study blast wave formation and propagation. 
	The shape, cavitation and position of the blast wave was characterised using an optical probe.
	%in preparation for full power shots. 
	%Unfortunately, we were unsuccessful in capturing on-shot probe images. 
	With the prepulse aligned to \SI{400}{\micro\metre} above the nozzle, the expected peak density after the formation of the blast wave was $8.9\,{n_c}$ since the unshaped gas jet was measured to have an electron density of $1.5\,{n_c}$ at a height of \SI{400}{\micro\metre} above the nozzle and for hydrogen, $\gamma=1.41$, so $\rho=5.88\, \rho_0$ \cite{Sedov1959}.

	An example shadowgraphy image is shown in figure \ref{fig:probe}, clearly showing the high-density blast wave. The main beam focal position was referenced on the optical probe cameras by positioning a thin wire at focus and imaging it using a high magnification objective lens. This focal position is indicated in figure \ref{fig:probe} by a white cross, and was measured to be \SI{61}{\micro\metre} behind the blast wave which is comparable to the Rayleigh range, $z_r=\SI{63}{\micro\metre}$. Images taken with the optical probe at different relative delays to the prepulse beam allowed the blast wave radius to be measured as a function of time, as shown in figure \ref{fig:blastwave_radius_v_time}. The radius was measured by manually fitting a semi-circle to the right-hand end of the blast-wave.
	In order to determine the gas profile from the experiment, we can refer to the analytical Sedov solution for blast wave expansion \cite{Dover2016a,Sedov1959}. The solution is self-similar, such that the blast wave radius $r_\mathrm{B}$ with dimensionality $\alpha$ expands with time $t$ as,

\begin{equation}
    r_\mathrm{B}(t) = \lambda^*\left(\frac{E_{ab}}{\rho}\right)^{1/\left(2+\alpha\right)} t^{2/\left(2+\alpha\right)},
    \label{eq:sedov}
\end{equation}
where $\lambda*$ is a constant of order unity, $E_{ab}$ is the deposited energy that initiated the blast wave with units $ML^{\alpha-1}T^2$ and $\rho=n_em_p$ is the mass density of the background plasma. At a height of \SI{400}{\micro m} above the nozzle $\rho=n_em_p=\SI{2.53}{\kilo\gram\per\metre\cubed}$. The dimensionality was taken to be $\alpha=3$, since the blast wave was observed to expand spherically at the end of the laser deposition region, despite a cylindrical expansion behind this point.  Using equation \ref{eq:sedov}, a best fit energy was found to be $E_{ab}=(31\pm{7})\,\si{\milli\joule}$ which corresponds to a deposition efficiency of 15$\pm{3}$\%.

The proton data presented in figure \ref{fig:TP_data}\,(e-g) was taken with a relative delay between the peak of the prepulse and the main pulse of 6.18\,ns. Using the fit in figure \ref{fig:blastwave_radius_v_time}, the blast wave radius, $r_\mathrm{B}(\SI{6.18}{\nano\second})=\SI{216}{\micro\metre}$. By mass conservation of a spherical shell of thickness $\delta r(t)$, with the initial plasma in the limit $\delta r\ll r_B$, the shell thickness of the blast wave can be shown to be \cite{Ettlinger2018}
\begin{equation}
\delta r(t) = \frac{\gamma-1}{3(\gamma+1)}r_\mathrm{B}(t).
\end{equation}
At the time of the main pulse interaction the shell thickness was $\delta r(\SI{6.18}{\nano\second})=\SI{12.4}{\micro\metre}.$
The rear density scale-length can be approximated to the ion mean free path.  The mean free path was calculated as the distance travelled by ions moving at the thermal velocity $v_i=\sqrt{{2k_B T_i}/{m_i}}$ in the ion collision time $\tau_i=6.6\times10^{-10}{\sqrt{A {T_{i,\mathrm{keV}}}^3}}/{(n_i{Z_i}^4ln{\Lambda_i})},$ in which $T_{i,\mathrm{keV}}$ is the ion temperature in keV, $A$ is the ion mass number, $n_i$ is the ion density in $10^{27}$\si{\per\metre\cubed}, $Z_i$ is the ionisation state and $\ln{\Lambda_i}$ is the Coulomb logarithm \cite{Atzeni2004}.
The ion temperature, $T_i$, in the post-shock region is given by $k_BT_i=2\left({(\gamma-1)}/{(\gamma+1)^2}\right)\left({Am_i {v_{sh}}^2}/{(Z+1)}\right)$ in which $v_{sh}$ is the shock velocity \cite{Drake2005}. $v_{sh}$ was calculated from the time derivative of equation \ref{eq:sedov} giving $v_{sh}(t=6.18\,\mathrm{ns})=$\SI{14.0}{\kilo\metre\per\second}, resulting in an ion temperature of \SI{0.14}{\electronvolt} and a mean free path \SI{0.65}{\pico\metre}. The resulting gas density profile is plotted in figure \ref{fig:gas_jet_profiles}, where we have taken the blast wave to be positioned at the centre of the gas jet.
	
The full power shots into shaped gas (fig. \ref{fig:TP_data}\,(e-g)) produced dramatically different results from those without shaping. The radial proton signal previously observed disappeared and forward going protons were detected. Thomson parabola image plate scans from one of the two shots into shaped gas which observed protons are shown in figures \ref{fig:TP_data}\,(e-g). Of particular interest is the proton bunch from the +\ang{5.9} TP shown in figure \ref{fig:TP_data}\,(f). The proton bunch is observed at \SI[separate-uncertainty = true]{1.32(2)}{MeV} with a \SI[separate-uncertainty = true]{12.9(3)}{\%} energy spread and charge \SI[separate-uncertainty = true]{400(30)}{\pico\coulomb\per\steradian}. The protons detected at \ang{-11.1} also had a charge of \SI[separate-uncertainty = true]{400(30)}{\pico\coulomb\per\steradian}. Again this narrow energy feature is indicative of shock acceleration \cite{Palmer2011, Haberberger2011}. Its direction suggests that the shock formation region is normal to the laser beam propagation. This is what would be expected if, in this instance, the plasma density has exceeded the critical density and the shock is generated at this surface.

	\section{Simulations}\label{sec:simulations}
	%=================================================================================
	Simulations were carried out for both the undisturbed gas jet and the optically shaped gas jet using the particle-in-cell (PIC) code SMILEI \cite{Derouillat2018}. The simulations used two spatial and three velocity dimensions (2D3V), a cell size of 30\,nm$\times$60\,nm, a simulation domain of 1.2\,mm$\times$0.25\,mm, and a timestep of 88.6\,as. The proton and electron particles were initiated cold with 4 particles per cell. The simulation was diagnosed on a grid with 120\,$\times$120\,nm cells. For the unshaped gas jet, the experimentally measured gas jet profile was used, with peak density 0.82\,$n_c$ as shown in figure \ref{fig:gas_jet_profiles} and the laser was focussed at the centre of the gas jet. For the optically shaped gas jet, the Sedov solution with peak density 8.9\,$n_c$ was used, which is also shown in figure \ref{fig:gas_jet_profiles}. The laser was focussed \SI{61}{\micro\metre} behind the front of the blast wave, as in the experiment. 
	
	\begin{figure}
	\centering
	\includegraphics[width=1\textwidth]{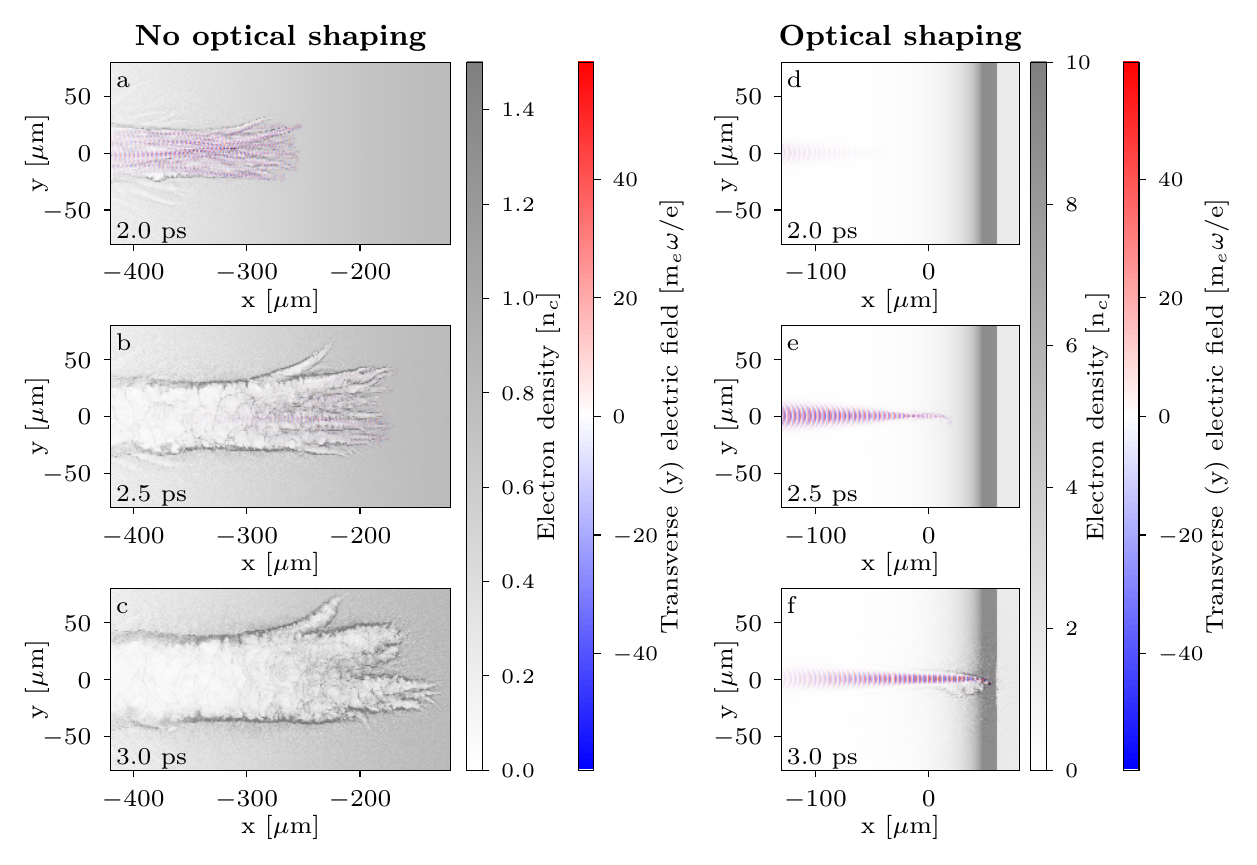}
		\caption[Simulation real space]{Electron density (grey) and transverse electric field (colour) spatial distributions from SMILEI PIC simulation of an undisturbed gas jet at: a) \SI{1.9}{\pico\second}, b) \SI{2.5}{\pico\second}, c) \SI{3.0}{\pico\second}; and of shaped gas jet at: d) \SI{1.9}{\pico\second}, e) \SI{2.5}{\pico\second}, f) \SI{3.0}{\pico\second}. The $x$-axis corresponds to the distance from the laser focal position. The electron density in f) reaches 64.2$n_c$, although the colour-scale has been truncated to show the lower-density structure.}
		\label{fig:Simulation_real_space}

	\end{figure}

	%Figure \ref{fig:Simulation_phase_space} shows the phase space for the momenta in $x$ and $y$, where $x$ is the laser propagation direction and $y$ is the transverse direction. Each pixel corresponds to the sum of macro-particles with momenta ${p_x}$ and ${p_y}$ according to the two axes. The higher proton numbers at the centre of the plot correspond the particles at very low energy. The further a pixel from the centre of the box the higher its corresponding energy. The 5 superimposed circles correspond to proton energies of 2, 10, 20, 30 and 50\,\si{\mega\electronvolt}. The solid and dashed black segments extended in positive ${p_x}$ and positive ${p_y}$ are the regions of phase space used when calculating the spectra for longitudinally and transversely accelerated protons respectively.
	
	Figure \ref{fig:Simulation_real_space} shows the transverse electric (i.e. primarily laser) field overlayed on top of the corresponding electron density from two different simulations, with and without optical shaping. Figure \ref{fig:Simulation_real_space}(a) shows the unshaped gas density profile at \SI{2.0}{\pico\second} into the simulation. The laser has filamented over a region $\approx \SI{50}{\micro\metre}$ transversely. By \SI{2.5}{\pico\second}, as shown in figure \ref{fig:Simulation_real_space}(b), the laser has started hosing and the laser energy is mostly depleted. By \SI{3.0}{\pico\second}, as shown in as shown in figure \ref{fig:Simulation_real_space}(c), the laser energy is depleted and the laser filamentation can be seen imprinted on the electron density. High-density laser-formed channel walls can be seen at $y=\pm \SI{40}{\micro\metre}$.
	The expansion of these walls through the colder background plasma at later times leads to shock acceleration of ions from this background at \ang{90} to the laser propagation.
	
	Figure \ref{fig:Simulation_real_space}(d) shows the electron density and transverse electric field from a simulation of the shaped gas density profile at \SI{2.0}{\pico\second}. Due to the cavitation behind the blast wave, the laser has propagated without instabilities. After \SI{2.5}{\pico\second}, as shown in figure \ref{fig:Simulation_real_space}(e), the laser has self-focussed at $x=\SI{-25}{\micro\metre}$ due to the low-density plasma in the rising edge of the Sedov structure. Upstream of the self-focussing, the laser has begun to filament. Figure \ref{fig:Simulation_real_space}(d) shows the same simulation at \SI{3.0}{\pico\second}. The laser has reached the overdense region of the gas and a high-density (64.2\,n$_c$) shock region has been generated at the relatistically-corrected critical density surface.
	
	\begin{comment}
	\begin{figure}
		\centering
		\includegraphics[width=1\textwidth]{images/shaped_v_unshaped_simulation_phase_space.pdf}
		\caption[Simulation phase space]{$p_xp_y$ phase space from a simulation of an a) unshaped gas jet with peak density $1.7\,n_c$ and b) a Sedov density profile with peak density $9.8\,n_c$. The black dashed arc corresponds to the angle of the +\ang{90} TP. The black solid arc corresponds to the full range of all 3 forward TPs. The 5 circles correspond to proton energies of 2, 10, 20, 30 and 50\,\si{\mega\electronvolt}}
		\label{fig:Simulation_phase_space}
	\end{figure}
	\end{comment}
	
	%Figure \ref{fig:Simulation_phase_space}a shows the phase space for the unshaped gas case. The distribution is elongated in the ${p_y}$ direction with protons of energies over \SI{35}{\mega\electronvolt}, whereas in the longitudinal direction, there are no protons over 10\,MeV. Figure \ref{fig:Simulation_phase_space}b shows the phase space for the constant electron density case. The transverse acceleration has been significantly reduced compared to the unshaped case, and there is increased longitudinal acceleration, with energies over \SI{20}{\mega\electronvolt}. The hole-boring velocity from this simulation is $v_{\mathrm{hb}}=0.087\mathrm{c}$. Protons accelerated by the hole-boring mechanism would have an energy corresponding to $v_{\mathrm{proton}}=2v_{\mathrm{hb}}$ which is $E_{\mathrm{proton}}=14.1\,\si{\mega\electronvolt}$.
		\begin{figure}
		\centering
		\includegraphics[width=12cm]{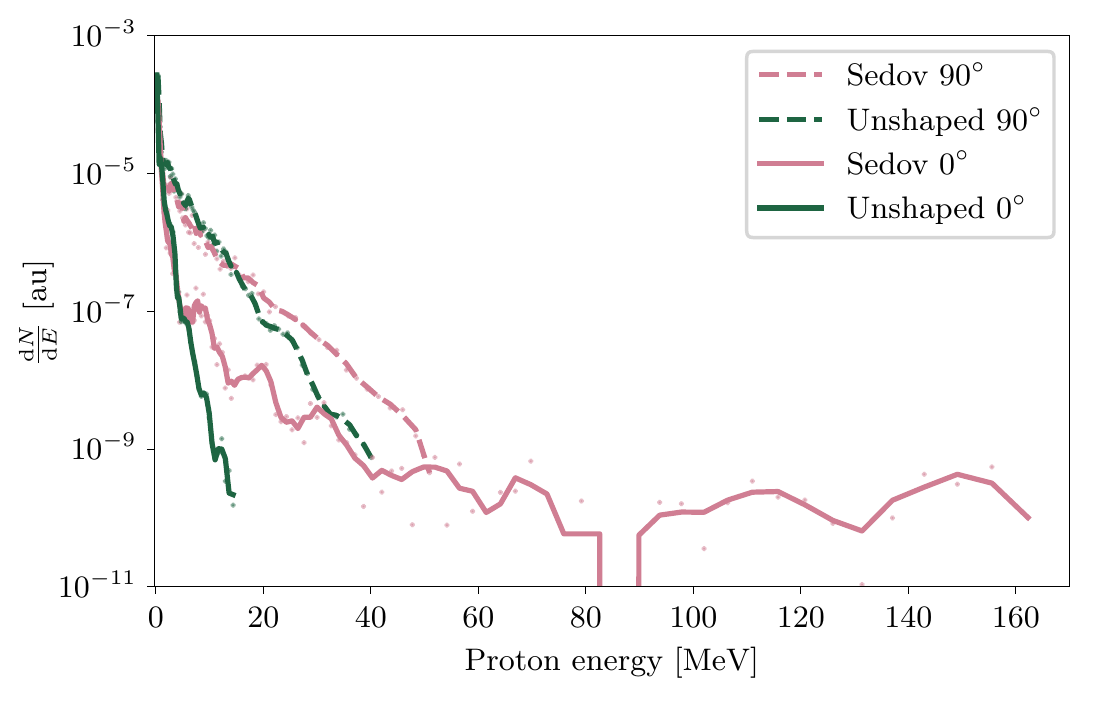}
		\caption[Simulation spectra]{Proton energy spectra from simulations for the unshaped gas jet (green) and for a shaped gas jet (pink). The solid lines correspond to protons accelerated in the laser propagation direction and the dashed lines correspond to protons accelerated perpendicular to the laser propagation direction. A 3 pixel moving average filter was applied to the spectra, the unfiltered individual data points are also plotted.}
		\label{fig:Simulation_spectra}
	\end{figure}
	Figure \ref{fig:Simulation_spectra} shows the extracted proton spectra from the PIC simulations. The solid lines are spectra intergrated over protons propagating within $\pm\ang{2.5}$ of the laser axis and the dashed lines are similarly intergrated over protons propagating at $\ang{90}\pm\ang{2.5}$ with respect to the laser axis. In pink are the spectra for the (shaped) Sedov gas profile and in green are the spectra for the unshaped gas jet. The energies in the simulation are much higher than measured in the experiment and no radially propagating protons were observed in the experiment. However, the simulations do reproduce key qualitative aspects of the experimental observations.
	The simulation with the unshaped gas jet has higher proton energies and a larger flux in the radial direction than the longitudinal direction.
	For the Sedov gas profile, which models the optically shaped target, proton energies extend to much higher values in the longitudinal direction. These protons have been accelerated by shock acceleration, as can be witnessed by the separation of this population from the lower energy protons. 	
	The relative dominance of radial to longitudinal acceleration for shaped and unshaped targets is as observed in the experiment.

	In the case of the unshaped profiles, the laser never reaches the critical surface, 
	The laser loses energy rapidly as it traverses the longer density scale-length, due to the generation of a ponderomotive channel, laser filamentation and laser hosing, as shown in figure \ref{fig:Simulation_real_space}(a).  As a result, the laser travels only in the underdense ramp of the gas jet, and never reaches a critical density surface to drive a forward-going shock.
	For the Sedov density profile case, the underdense region is significantly reduced and the laser reaches the critical density surface, launching a collisionless shock that promotes longitudinal acceleration. 
	
	% sort out times if figures
	%=================================================================================
	\section{Discussion}\label{sec:conclusion}
	%=================================================================================
	We have observed high energy protons from a high pressure gas jet driven by the Vulcan Petawatt laser. 
	Without optical shaping no forward propagating protons were detected. However, a thermal spectrum was detected perpendicular to the laser propagation direction with a high energy bunch at \SI[separate-uncertainty = true]{8.27(7)}{MeV}. With optical shaping of the density profile, forward going protons were detected. These protons had an energy of \SI[separate-uncertainty = true]{1.32(2)}{MeV} but were bunched with a \SI[separate-uncertainty = true]{12.9(3)}{\%} energy spread. 
	%The suppression of radial acceleration and enhancement of longitudinal acceleration is supported by particle-in-cell simulations, as shown in figure \ref{fig:Simulation_spectra}. 
	
\begin{figure}
		\centering
		\includegraphics[width=0.4\textwidth]{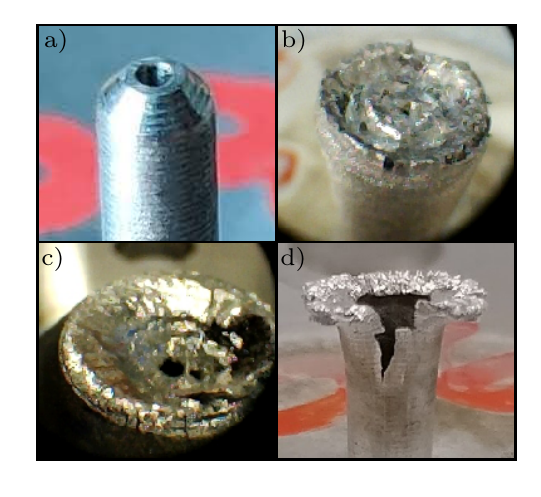}
		\caption[nozzles]{Images of a nozzle a) before a full power shot and b-d) after a single full power shot with each nozzle. In b) and c) the top of each nozzle has been melted by the interaction and in d) the nozzle has been forced open.}
		\label{fig:nozzle_damage}
	\end{figure}

	However, the observed proton energy of \SI[separate-uncertainty = true]{1.32(2)}{MeV} is low considering the expected laser intensity of $I_{\mathrm{MP}}=7.4\times10^{24}\,\si{Wm^{-2}}$ and the \textgreater 100\,MeV protons observed in PIC simulations. Additional PIC simulations with an increased laser spot-size showed that a reduction of the laser intensity \textit{on-target} by a factor of $\sim64$ supports these proton energies. We also note that 2D simulations do not fully describe many of the processes such as laser filamentation, and so are also likely to lead to exaggerated ion energies. Due to the very large computation requirements, full 3D simulations will be the subject of future work.
	Another technical challenge for future experiments is displayed in Figure \ref{fig:nozzle_damage}. Figure \ref{fig:nozzle_damage}(a) shows a nozzle before any full power shots. Figure \ref{fig:nozzle_damage}(b-d) shows images of 3 different gas nozzles after they had each been used on a single full power shot. The nozzles show considerable damage. Full power shots with no gas, gas with no laser and gas with low power shots resulted in no damage. This suggests laser energy passes to the nozzle via the gas and melts the tip of the nozzle. During the interaction electrons are expelled from the plasma much faster than the protons due to their lower mass. This results in a strong return current through the nozzle into the plasma that can resistively heat the nozzle until it melts. Most of the nozzles used were made from aluminium, but steel and gold-coated steel nozzles were also used and equally damaged.
	
	This resistive heating of target structures has also been observed in solid target experiments \cite{Bradford2018}. 
	The damaged nozzles cause turbulence in the gas flow and thus reduces the repeatability of shots. As a result, in this work the nozzle was changed on each shot.
	In order to use this acceleration scheme at high repetition-rates, a solution to this nozzle damage problem will need to be found such as moving the nozzle further below the interaction, the use of sacrificial parts to protect the nozzle surface or by manufacturing the nozzle from an insulating material.

	Nevertheless, we have been able to demonstrate that it is possible to produce short enough scale-length high-density gas plasmas to investigate shock acceleration with optical or near-infrared lasers. With better control of laser parameters such as a co-propagating prepulse to preform the chanel and reduce filamentation, and overcoming target damage, this method is a promising route to producing compact sources of energetic ions.

\section{Acknowledgements}
%=================================================================================
We thank the staff of Vulcan at the Central Laser Facility, Rutherford-Appleton Laboratory for their support. We acknowledge funding from the following grants EPSRC EP/K022415/1, EPSRC/DSTL EP/N018680/1 and STFC ST/P002021/1. \\

	\bibliography{bibfile}
\end{document}